\documentclass[reprint,pre,amsmath,amssymb,aps,floatfix,superscriptaddress,longbibliography]{revtex4-2}

\usepackage[english]{babel}
\usepackage[utf8]{inputenc}
\usepackage[colorinlistoftodos, color=green!40, prependcaption]{todonotes}
\usepackage{amsthm}
\usepackage{hyperref}
\usepackage{amsmath}
\usepackage{amssymb}
\usepackage{mathtools}
\usepackage{physics}
\usepackage{xcolor}
\usepackage{graphicx}
\usepackage[left=20mm,right=20mm,top=25mm,columnsep=15pt]{geometry} 
\usepackage{comment}
\usepackage{adjustbox}
\usepackage{placeins}
\usepackage[T1]{fontenc}
\usepackage{lipsum}
\usepackage{csquotes}
\newcommand{\tg}[1]{\textcolor{blue}{#1}}
\newcommand{\lh}[1]{\textcolor{brown}{#1}}
\newcommand{\highlightsname}{Research highlights}
\newenvironment{highlights}{\par\noindent\textbf{\highlightsname}\par\begin{itemize}}{\end{itemize}}

\begin{document}
\title{Hierarchical organization governs nonlinear mechanical reversibility \\in $\iota$-carrageenan gels}



\author{Thomas Gibaud}
\email[]{Corresponding author, thomas.gibaud@ens-lyon.fr}
\affiliation{Univ Lyon, Ens de Lyon, CNRS, Laboratoire de Physique, 69342 Lyon, France}
\affiliation{Department of Polymer Engineering, IPC,University of Minho, Guimarães, 4804-533 Portugal}

\author{Loïc Hilliou}
\email[]{Corresponding author, loic@dep.uminho.pt}
\affiliation{Department of Polymer Engineering, IPC,University of Minho, Guimarães, 4804-533 Portugal}

\date{\today}

\begin{abstract}
Carrageenan gels are thermoreversible polysaccharide networks whose mechanical properties emerge from ion-mediated helix association, yet how their molecular organization controls nonlinear deformation remains poorly understood. Here, we investigate the temperature-dependent rheology of $\iota$-carrageenan gels formed in KCl solutions using linear and nonlinear oscillatory rheology combined with normal-force measurements.
$\iota$-carrageenan forms homogeneous and mechanically reversible gels whose elastic modulus increases continuously with quench depth. Deep quenches generate pronounced strain stiffening before yielding, associated with the development of internal stresses revealed by negative normal forces. Remarkably, large deformations preserve the small-strain elastic modulus while progressively suppressing strain stiffening, demonstrating a partial mechanical reversibility of the network.
We interpret these observations using a hierarchical network picture in which a persistent intermolecular network controls linear elasticity, while a more fragile mesoscale organization enables cooperative alignment and stress amplification under deformation. Our results highlight that nonlinear mechanics of thermoreversible polysaccharide gels are governed not only by molecular connectivity, but also by the reversible formation and eventually destruction of mechanically adaptive hierarchical structures.
\end{abstract}


\maketitle

\section{Introduction}

Polysaccharide gels are among the most widely studied soft materials, both for their practical relevance as food hydrocolloids, pharmaceutical excipients, and biomedical scaffolds, and for their fundamental interest as model systems in which network formation, viscoelasticity, and nonlinear mechanics can be tuned by molecular architecture and ionic environment~\cite{clark1987}. Beyond their ability to form gels, {polysaccharides} are valued for their capacity to provide specific textural attributes, including firmness, elasticity, fracture resistance, and recovery after deformation. These properties are governed not only by the small-strain elastic modulus {of the gel network}, but also by nonlinear mechanical responses such as strain stiffening, yielding, and structural recovery after large deformations, which determine how gels deform during {fabrication} and oral {processing}~\cite{foegeding2007,smirnov2024}. A central challenge is therefore to understand how molecular self-assembly gives rise to {a network displaying} these distinct mechanical functions. Recent advances in polymer and hydrogel science have shown that hierarchical architectures, such as double-network hydrogels and sacrificial bond networks, can decouple elasticity from nonlinear reinforcement by assigning load-bearing and energy-dissipating functions to different structural elements~\cite{gong2003}. Whether naturally occurring thermoreversible polysaccharides achieve a similar functional separation through their intrinsic self-assembly remains largely unexplored.

Among thermoreversible polysaccharides, carrageenans provide an ideal model system to investigate how molecular assembly pathways translate into nonlinear mechanical functions. Extracted from red seaweeds, carrageenans form thermoreversible gels upon cooling in the presence of monovalent or divalent cations, and their gelation mechanism is among the best characterized of all polysaccharide systems at the molecular level~\cite{smidsrod1982,morris1980}. They are linear sulfated polysaccharides whose two main gelling forms, $\iota$- and $\kappa$-carrageenan, differ by a single sulfate group per disaccharide unit. 
This subtle chemical difference strongly affects the organization of the resulting networks. Upon cooling in KCl solutions, both systems undergo a coil-to-helix transition followed by cation-mediated association of helices into junction zones~\cite{morris1980,rochas1984}. In $\kappa$-carrageenan, K$^+$ ions promote extensive aggregation of helices into rigid bundle-like junction zones, leading to strong, turbid, and brittle gels that may exhibit syneresis and a transient overshoot of the elastic modulus during gelation~\cite{hermansson1991,yuguchi2002,takemasa2001}. In contrast, the additional sulfate group of $\iota$-carrageenan limits inter-helix aggregation through electrostatic and steric effects, resulting in a more hydrated, loosely connected, and dynamically rearranging network~\cite{smidsrod1982,van2002,takemasa2001}. Consequently, $\iota$-carrageenan forms transparent and more deformable gels that display pronounced strain stiffening before yielding~\cite{morris1980,hilliou2021,yang2020}. Although the molecular basis of carrageenan gelation is well established, a quantitative physical description connecting molecular organization to nonlinear mechanical response remains elusive~\cite{hilliou2021}. 

Classical structural models successfully describe the coil--helix transition, ion-specific stabilization, and helix aggregation into junction zones~\cite{morris1980,smidsrod1982,rochas1984,yuguchi2002}, but they do not explicitly account for the generation, storage, and redistribution of mechanical stresses during network formation. In particular, the origin of strain stiffening in $\iota$-carrageenan remains unclear, as it has not yet been linked to a specific structural mechanism of deformation. More generally, the role of internally generated stresses during gelation---arising from network contraction, syneresis, tensegrity-like force balance, and topological frustration---has not been incorporated into current descriptions of carrageenan mechanics, despite growing evidence that such residual stresses strongly influence the rheology of thermoreversible polysaccharide gels~\cite{mao2016,wu2024} and fibrous networks~\cite{Meng2016}. Finally, while temperature controls the coil--helix transition, it remains unclear whether the depth of cooling simply modifies the density of intermolecular constraints or instead changes the formation pathway and mechanical organization of the network, thereby controlling its nonlinear response.


Here, we address these questions through a systematic rheological study of $\iota$-carrageenan gels formed in KCl solutions, combining linear and nonlinear oscillatory rheology, temperature-dependent gelation monitoring, and normal force controlled experiments. Both normal force control during gelling and monitoring during nonlinear rheometry show novelty in the carrageenan literature. We show that $\iota$-carrageenan forms homogeneous and mechanically reversible gels governed by a connectivity-controlled gelation mechanism, in which elasticity increases continuously with quench depth following a critical-like scaling and shows pronounced strain stiffening. The mechanical reversibility of the gel's linear elastic modulus after stiffening and eventual yielding is rationalized by a theory describing the elasticity of rod-like filaments in the condensed state~\cite{doi1980}. The vanishing of the gel's strain stiffening after a first yielding is associated with a loss of tensegrity in the network~\cite{Meng2016} which is inferred by the recording of normal forces during the successive strain sweeps. However, both reversibility in the gel's linear elastic modulus and the progressive loss of strain stiffening upon repeated strain sweeps, cannot be explained by the theoretical descriptions of filamentous networks. Instead, a two-level structure is proposed to describe the mechanical behavior of $\iota$-carrageenan gels.  
These results open new directions for understanding the role of stress generation in thermoreversible biopolymer networks.

\section{Materials and methods}

\paragraph{Carrageenan solutions --}
$\iota$-carrageenan (Sigma C1138-
100G) was dispersed in water containing 100~mM KCl at a weight concentration of $c_w = 10$~mg/mL. The suspensions were placed in a hermetic container, stirred, and heated at $80^\circ\mathrm{C}$ for approximately 1~h until they became translucent.
The obtained carrageenan solutions were then loaded hot into the rheometer at $T = 80^\circ\mathrm{C}$.

The polymer concentration $c_w = 10$~mg/mL and the salt KCl at 100~mM were chosen to achieve a gel elasticity which remains in a range accessible by standard oscillatory rheology, whereas small variations in ionic strength associated with the presence of other salts in the commercial carrageenan used here, but in much smaller amounts, are screened by the added KCl.

\vspace{3mm}

\paragraph{Rheology protocols -- }
The rheological protocols are summarized in Tab.~\ref{tab:rheology_protocol}. Following sample loading and preshear to homogenize the thermal and mechanical history of the dispersion, the sample was equilibrated at high temperature in the sol state. Gelation kinetics were then monitored during controlled cooling down to the final temperature $T$, allowing the evolution of the viscoelastic properties during network formation to be followed continuously. After gelation, an additional equilibration step was performed to ensure that the gel reached a stable mechanical state prior to further characterization (steps~1--4).

\begin{table}
\centering
\begin{tabular}{|c|p{7cm}|}
\hline
Step & Description \\
\hline
\hline
1 & Set temperature to $T = 80^{\circ}$C, load sample and preshear for 60~s at shear rate $\dot{\gamma} = 10~$s$^{-1}$ \\
\hline
2 & Equilibration  at $T = 80^{\circ}$C: time sweep - measure $G'$ and $G''$ at $f = 1$~Hz and $\gamma = 3\%$ for 600~s \\
\hline
3 & Gelation kinetics: time sweep from $T = 80^{\circ}$C to $T$ - Measure $G'$ and $G''$ at $f = 1$~Hz and $\gamma = 3\%$ while performing a linear temperature ramp from $T = 85^{\circ}$C to $T$ at a rate of 0.033$^{\circ}$C/s \\
\hline
4 & Equilibration at $T$: time sweep -  measure $G'$ and $G''$ at $f = 1$~Hz and $\gamma = 3\%$ for 600~s (gelation monitoring) \\
\hline\hline
5a & Viscoelastic spectrum of the gel: frequency sweep at $T$ - perform a frequency sweep $f = 20 \rightarrow 0.002$~Hz with 7 points per decade at $\gamma = 3\%$ \\
\hline
6a & Equilibration at $T$ again: time sweep - measure $G'$ and $G''$ at $f = 1$~Hz and $\gamma = 3\%$ for 60~s (post-frequency sweep check) \\
\hline
7a & Yielding properties of gel: strain sweep at $T$ - perform a strain sweep from $\gamma = 1 \rightarrow 1000\%$ with 20 points per decade at $f = 1$~Hz \\
\hline
\hline
5b & Yielding reversibility properties of the gel: strain sweep at $T$ - perform a strain sweep from $\gamma = 0.1\rightarrow 2; 0.1\rightarrow5; ... ;0.1\rightarrow 1000\%$ with 20 points per decade at $f = 1$~Hz. After the i-1 strain sweep iteration at a maximum amplitude $\gamma_{max,i-1}$, we note $G_{0,i}$ the recovered elastic modulus at low strain and $F_0$ the associated normal force in plate-plate geometry. \\
\hline
\end{tabular}
\caption{Rheology measurement protocol.}
\label{tab:rheology_protocol}
\end{table}

Two complementary rheological characterizations were then performed on the fully formed gels. In the first protocol (steps~5a--7a), the linear viscoelastic properties were characterized by frequency sweeps, followed by strain sweeps to probe the nonlinear response and yielding behavior of the gels. In the second protocol (step~5b), successive strain sweeps of increasing amplitude were performed to investigate yielding reversibility and the structural recovery of the gel network after deformation alike in~\cite{moraes2024}.

The rheological measurements were carried out either using an Anton Paar MCR302 rheometer equipped with a plate--plate geometry (radius $R=2$~cm, gap $h=0.5$~mm), allowing the use of normal force control ($F_N$), or using an MCR300 rheometer equipped with a Couette geometry (outer radius 10~mm, gap 0.5~mm). The Couette is ideal to make sucessive strain sweeps without spilling the gel through the gap as compared to plate plate geometry. In both cases, light mineral oil (Sigma-Aldrich, ref. 330779) was applied around the sample to prevent evaporation during the measurements.

Normal-force-controlled rheology has emerged as an important approach to investigate gelation in systems undergoing volume changes, syneresis~\cite{leocmach2015,wu2023}, or internal stress build-up during network formation. In conventional fixed-gap measurements, thermoreversible polysaccharide gels may develop significant normal stresses as the sample contracts during gelation, leading to drift in the measured elastic modulus, debonding from the geometry, or mechanically constrained network formation. Mao \textit{et al.} showed for agar gels that imposing a constant normal force ($F_N=0$~N) instead of a fixed gap compensates for sample contraction and suppresses these artifacts, allowing reliable measurements of gelation dynamics and final elastic properties~\cite{mao2016,andrieux2023}.
Beyond correcting experimental artifacts, residual stresses are now recognized as an intrinsic feature of many arrested soft materials, where they govern aging, yielding, slow structural relaxation, and mechanical memory~\cite{cipelletti2000,wu2024,pommella2021}. Normal-force-controlled rheology therefore provides a unique opportunity to distinguish the intrinsic evolution of the gel network from stresses generated by mechanical constraints during gelation, enabling direct investigation of the coupling between network assembly and residual stress development.

\section{Results}

The rheological behavior of $\iota$-carrageenan gels is first analyzed as a function of the final gelation temperature to establish how cooling controls network formation and linear elasticity. We then investigate how the resulting networks respond to large deformations through strain stiffening, yielding, and mechanical reversibility.


\vspace{3mm}
\paragraph{Temperature sensitivity of $\iota$-carrageenan gels}

\begin{figure}
\centering
\includegraphics[width=0.48\textwidth]{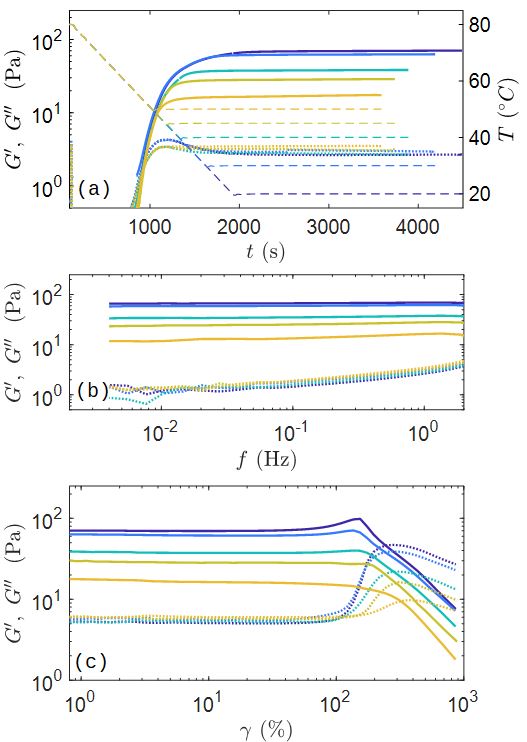}
\caption{Influence of the final temperature $T$ on $\iota$-carrageenan gels.
(a) Time evolution of the elastic ($G'$) and viscous ($G''$) moduli during a temperature quench from 80~$^\circ$C to $T = 20^\circ$C (blue), 30$^\circ$C (light blue), 40$^\circ$C (cyan), 45$^\circ$C (green), and 50$^\circ$C (yellow) with a cooling rate $v_T = 0.033^\circ$C/s.
(b) Corresponding viscoelastic spectra as a function of frequency $f$ at $T$.
(c) Yielding behavior as a function of strain $\gamma$.}
\label{fig:iota1_T}
\end{figure}

\begin{figure}
\centering
\includegraphics[width=0.48\textwidth]{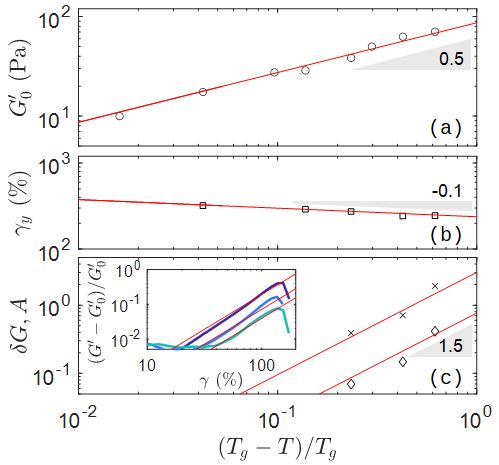}
\caption{Critical scaling of $\iota$-carrageenan gels  with respect to the reduce temperature $\mid T_g-T \mid/T_g$.
(a) Scaling of the gel linear elasticiy $G'_0$ after 2000~s. The red line is a power low fit with exponent 0.5.
(b) Scaling of the gel yield strain. The red line is a power low fit with exponent -0.1.
(c) Scaling of $\delta G=(G'_{max}-G'_0)/G'_0$ and  $A$. The red line is a power low fit with exponent 1.5.
Inset: $(G'(\gamma)-G'_0)/G'_0$ versus $\gamma$ at $T=20$, 30 and 40~$^{\circ}$C. Red lines are fit to the data with equation $y=A\gamma^2$.
}
\label{fig:Iota1_scale}
\end{figure}

As shown in Fig.~\ref{fig:iota1_T}(a), the gelation dynamics of $\iota$-carrageenan gels proceed smoothly when quenched below the gelation temperature $T_g=54$~$^{\circ}$C: the elastic modulus increases monotonically and saturate after roughly 2000~s for all tested $T$. 
The gel linear elasticiy $G'_0$ after 2000~s increases as the temperature quench gets deeper. As shown in Fig.~\ref{fig:Iota1_scale}(a), $G'_0$ follows a critical-like scaling with the distance to the gelation temperature, $
G'_0 \propto \mid T_g-T \mid ^{\alpha}$ 
with an exponent $\alpha \simeq 0.5$. 

\vspace{3mm}
\paragraph{Consequences on the viscoelastic spectrum --}
As shown in Fig.~\ref{fig:iota1_T}(b), The viscoelastic spectra remain qualitatively similar over the entire temperature range. The elastic modulus $G'$ is nearly independent of frequency, consistent with a percolated elastic network. 
The viscous modulus $G''$ is identical for all gels: it features a shallow minimum  around $f \simeq 0.01$~Hz. This low-frequency feature suggests the emergence of slow dissipative rearrangements. Since the minimum of $G''$ is temperature-independent, these gels do not follow the time–temperature superposition principle~\cite{williams1955, morris1984}. 

\vspace{3mm}
\paragraph{Consequences on yielding --}
As shown in Fig.~\ref{fig:iota1_T}(c), the nonlinear response evolves markedly with temperature. Close to the gelation temperature, the gels display strain softening and very large yield strains $\gamma_y$ defined when $G'=G''$, reaching $\gamma_y \simeq 350\%$, indicative of highly deformable and weakly constrained networks.
As the quench depth increases, the yield strain decreases slightly (see Fig.~\ref{fig:Iota1_scale}(b)), but the most prominent change is the emergence of pronounced strain stiffening. Above the onset of the nonlinear regime ($\gamma \gtrsim 10\%$), the elastic modulus increases with strain up to $G'_{max}$ before final yielding.
More precisely, for $\gamma \gtrsim 10\%$, the normalized elasticity up to $G'_{max}$ scales as  $(G'(\gamma)-G'_0)/G'_0 = A \gamma^2$ (Fig.~\ref{fig:Iota1_scale}(c-inset)). The prefactor $A$ varies with temperature.
As shown in Fig.~\ref{fig:Iota1_scale}(b), the relative stiffening amplitude scales as $\frac{G'_{\mathrm{max}}-G'_0}{G'_0}\propto\mid T_g-T\mid^{\beta_2}$ with $\beta_2\simeq1.5$.
The strain stiffening has a strong temperature dependence in contrast to the weak temperature dependence of $\gamma_y$. Indeed, the yield strain follows a critical scaling $\gamma_y \propto \mid T_g-T \mid^{-\beta_1}$ with $\beta_1\simeq 0.1$.
%

\begin{figure}
\centering
\includegraphics[width=0.48\textwidth]{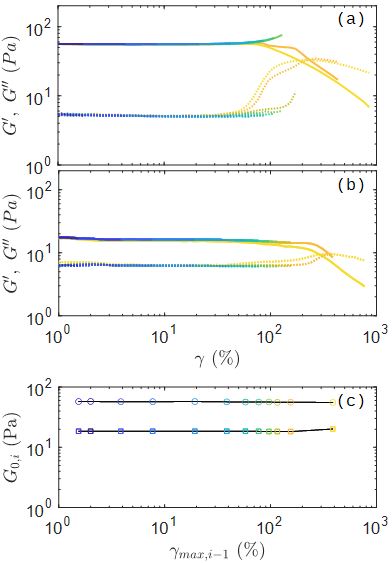}
\caption{Strain reversibility of $\iota$-carrageenan gels.
(a) Strain reversibility for gels formed at $T = 20~^\circ$C.The color code strain sweep with increasing maximum strain from blue to yellow. 
(b) Strain reversibility for gels formed at $T = 50~^\circ$C.
(c) Evolution of $G'_{0,i}$ (the linear elastic modulus measured during the strain sweep i) as a function of $\gamma_{\mathrm{max},i-1}$ (the maximum strain achieved during the strain sweep i-1) for gels formed at $T = 20~^\circ$C (circles) and $T = 50~^\circ$C (squares).}
\label{fig:Iota1_revers}
\end{figure}

\paragraph{Consequences on strain reversibility --}
Successive strain sweep experiments reveal that the nonlinear response remains reversible up to strains corresponding approximately to $G'_{\mathrm{max}}$ when the gel displays strain stiffening for deep quenches (Fig~\ref{fig:Iota1_revers}(a)) or up to strains of 110\% when the gel shows strain softening for shallow quenches (Fig~\ref{fig:Iota1_revers}(b)).
We note for both cases that the gel fully recovers its initial elastic modulus $G'_0$, demonstrating efficient structural recovery and network reorganization after deformation (Fig~\ref{fig:Iota1_revers}(c)). However, the following strain sweep after reaching $G'_{\mathrm{max}}$ for deep quenched gels, or a strain of 110\% for shallow quenches, displays enhanced softening. In particular, the strain-stiffening response disappears completely for the deep-quenched gel. This indicates that the mechanisms responsible for nonlinear stiffening are selectively altered by large deformations, while the underlying load-bearing network responsible for linear elasticity remains preserved. 


\section{Discussion}

The results reveal that the mechanical behavior of $\iota$-carrageenan gels cannot be understood solely from their linear viscoelastic properties. {The decoupling between linear elasticity reversibility and nonlinear irreversibility suggests that $\iota$-carrageenan gels contain mechanically distinct structural levels: a persistent network responsible for linear elasticity and a more strain sensitive connected organization responsible for strain stiffening and softening}. In this section, we first discuss how temperature controls network formation and elasticity, then examine the origin of strain stiffening and internal stresses, and finally show how the coexistence of reversible elasticity and irreversible nonlinear remodeling points to a hierarchical organization of the gel network.

\vspace{3mm}
\paragraph{Connectivity-controlled formation of a reversible $\iota$-carrageenan network -- }  
$\iota$-carrageenan gels display a continuous and homogeneous strengthening of the network upon cooling. The resulting elastic plateau modulus follows a critical-like scaling with the quench depth, $G'_0 \propto |T_g-T|^{\alpha}$, with an exponent $\alpha \simeq 0.5$. Such scaling behavior is characteristic of gelation processes governed by the progressive increase of network connectivity and intermolecular associations upon cooling. Similar power-law evolutions of the elastic modulus near thermoreversible gel transitions have been reported in polysaccharide and biopolymer systems, where elasticity emerges from the gradual percolation of transient junction zones~\cite{winter1986,chambon1987,norton1986}.
The relatively small exponent ($\alpha \simeq 0.5$) suggests that the increase in elasticity is not dominated by the formation of rigid aggregated bundles, but rather by a smooth reinforcement of a weakly connected and dynamically rearranging network. This interpretation is consistent with structural studies showing that $\iota$-carrageenan forms loosely associated helices with limited inter-helix aggregation due to its higher sulfate density~\cite{morris1980,van2002,yuguchi2002}. Indeed, the self-assembly of iota-carrageenan can be transposed from the concepts of hierarchical structures of proteins: intramolecular single helices (secondary structures) can coil into intramolecular super-helices (tertiary structures) which eventually coil in intermolecular super-helices (quaternary structures)~\cite{diener2019}. In this framework, cooling continuously increases both the density and lifetime of intra- and intermolecular constraints, leading to a progressive increase in elasticity while preserving homogeneous stress redistribution throughout the gel.

\begin{figure}
\centering
\includegraphics[width=0.48\textwidth]{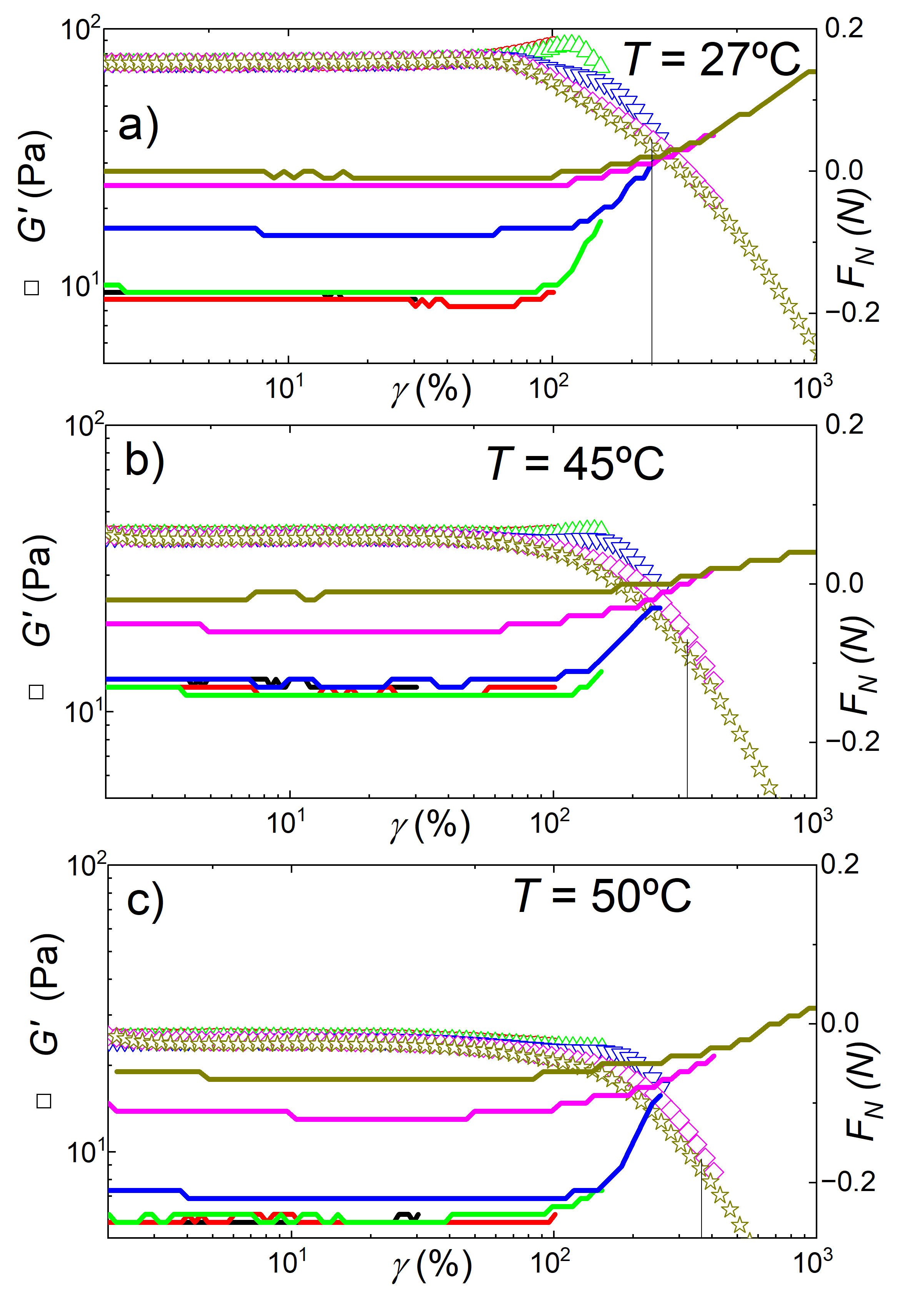}
\caption{$\iota$-carrageenan gels formed under zero normal force.
(a) Strain reversibility for gels formed at $T = 27~^\circ$C. The color code strain sweep with increasing maximum strain from black to red, green, blue, purple and dark yellow. 
(b) Strain reversibility for gels formed at $T = 45~^\circ$C.
(c) Strain reversibility for gels formed at $T = 50~^\circ$C. Vertical lines indicate the yield strain $\gamma_y$}.
\label{fig:Tf}
\end{figure}

\vspace{3mm}
\paragraph{Strain stiffening as a signature of mechanically adaptive network structures -- }
The nonlinear rheology further supports this picture. At deep quenches, the gels exhibit pronounced strain stiffening prior to yielding, indicating that deformation progressively aligns and tensions correlated network segments before failure, similarly to semiflexible or transient polymer networks~\cite{storm2005,licup2015,Meng2016}.  Theories for strain stiffening networks of semiflexible filaments essentially rely on the non linear deformation of a single filament. Molecular parameters such as the stiffness of the filament, filament's final stretchability and the network tensegrity originating from the crosslinking of filaments into the network, are at the core of strain stiffening. Interestingly enough, the tensegrity leads to the occurrence of negative normal forces under strain, that is, a negative Poynting effect~\cite{Meng2016}. Therefore, strain sweeps in parallel plates geometry were conducted in view to monitor the strain-induced normal forces $F_N$ during the strain stiffening. The results for a gel formed after a deep quench ($T = 27~^\circ$C) under zero normal force~\cite{mao2016} are reported in Fig~\ref{fig:Tf}(a). In the linear and strain stiffening regimes, a negative $F_N$ is measured which grows more negative, before reverting towards positive values as the maximum gel elasticity is passed by. This negative $F_N$ is in harmony with the existence of internal stresses emerging during the gel formation and giving rise to the filaments' tensegrity. The agreement with theoretical predictions~\cite{Meng2016} even encompasses the progressive vanishing of negative normal forces for gels formed closer to Tg (see Fig~\ref{fig:Tf}(b)), which still show strain stiffening. However, for $T = 50~^\circ$C, the negative normal force resulting from the gel formation does not originate any strain stiffening, see Fig~\ref{fig:Tf}(c). Within the same theoretical framework, this can be explained by a thermally-induced reduction of the filaments' stiffness, whereas tensegrity in the network is maintained. 

The strain reversibility of gels formed in the parallel plate geometry is again evidenced in the strain dependence of the dynamic shear moduli in Fig~\ref{fig:Tf}. The linear elastic moduli are recovered even after a strain sweep up to  $\gamma_{\mathrm{max},i-1}$ > $\gamma_y$. Strain stiffening is fully recovered, provided that $\gamma_{\mathrm{max},i-1}$ does not exceed the strain where $G'_{max}$ is reached. However, beyond $G'_{max}$, $F_N$ does not show mechanical reversibility, as the normal force keeps memory of its value reached at $\gamma_{\mathrm{max},i-1}$. As a result, following strain sweeps show less strain stiffening since the network has released some internal stresses upon the previous sweeps, and there is less tensegrity on filaments to promote stiffening. Eventually, strain stiffening vanishes when $F_N$ goes positive after the full yielding of the gel, see Fig~\ref{fig:Tf}(a) and (b). 

Within the elastic theory for semiflexible filament networks, the loss of network tensegrity inferred from the strain evolution of $F_N$ during repeated strain sweeps is not compatible with the fact that the linear elastic modulus is fully strain reversible. Meng and Terentjev~\cite{Meng2016} showed that $G_0$ is a function of both the tensegrity and the filaments' stiffness. Therefore, if $F_N$ becomes more positive after repeating strain sweeps, $G_0$ should drop. Alternatively, strain stiffening in networks of semiflexible contacting filaments can be simply rationalized by the local strain-induced positional rearrangement of neighboring thick filaments, which then bend and stretch the other contacting filaments in the network~\cite{doi1980}. In this theoretical framework, mechanical reversibility stems from the absence of crosslinks, since simple contacts between filaments crowding the volume are at the origin of the network´s elasticity $G_0$, whereas the bending-stretching and full shear orientation of filaments are responsible for the strain stiffening and eventual yielding. Moreover, Doi and Kuzuu~\cite{doi1980} showed that $G_0$ does not depend on the length of the filament. As such, elastic reversibility beyond $\gamma_y$ can be achieved even when shear-induced break-up of filaments occurs. However, this theory does not account for the disappearance of strain stiffening after large deformations, despite complete recovery of the small-strain modulus $G'_0$.

\begin{figure*}
\centering
\includegraphics[width=1\textwidth]{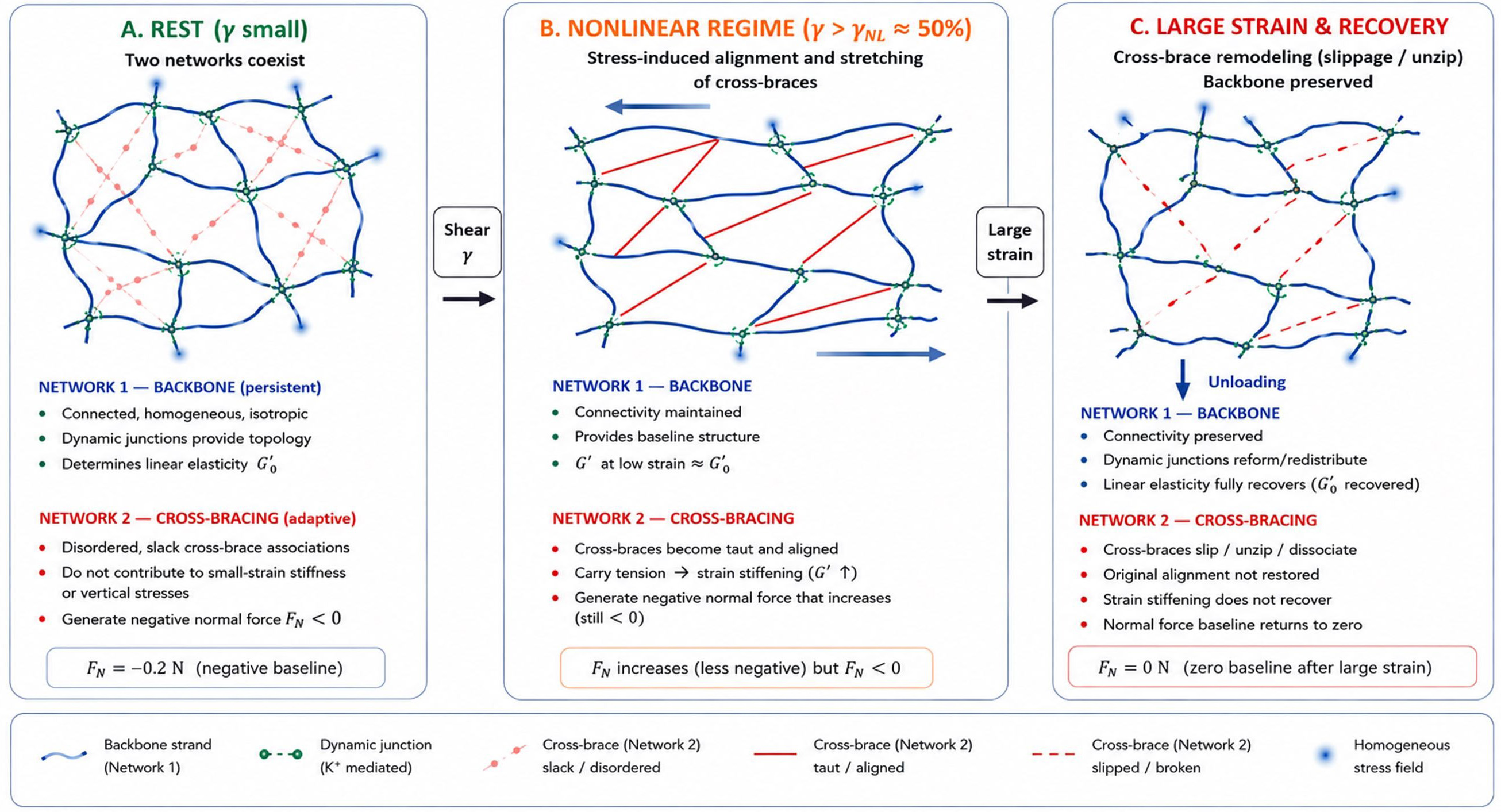}
    \caption{sketch of the hierarchical dual adaptive structure of $\iota$-carrageenan gel according to the \textit{Beach Sling Chair} model which is responsible for the strain reversible linear elastic modulus (Network 1 in blue) and the irreversible strain stiffening (Network 2 in red).}
        \label{fig:sketchIota}
\end{figure*}

\vspace{3mm}
\paragraph{Hierarchical organization enables reversible elasticity and nonlinear mechanical memory -- }
Neither of {the above theories} can account for all experimental observations. The filament-contact model explains the recovery of the linear modulus but does not predict the irreversible loss of strain stiffening. Conversely, tensegrity-based models naturally account for the normal-force signature but imply that loss of internal stress should also modify the linear elasticity. The coexistence of reversible {linear} elasticity and irreversible nonlinear response therefore requires an additional structural degree of freedom.
To resolve this apparent contradiction, we propose that strain stiffening originates from a hierarchical network organization involving two mechanically distinct structural levels, schematized in Fig.~\ref{fig:sketchIota}, i.e. the \textit{Beach Sling Chair} model.

Here, we propose a hierarchical network in which a {strain-}persistent percolated backbone governs linear elasticity, while a more fragile mesoscale organization is responsible for nonlinear strain stiffening. The second level may correspond to mesoscale associations of helical domains or cooperative aggregates of intermolecular helices {(the quaternary structures proposed for carrageenans \cite{diener2019})} formed during {deeper} cooling. Large deformations likely disrupt part of these inter-helix associations or orientational correlations that initially enabled cooperative alignment and tension transfer under deformation. The gel therefore recovers its underlying connectivity at small-strain elasticity, but no longer contains sufficiently extended correlated structures capable of further alignment and strain-induced stiffening. The absence of strain stiffening close to $T_g$ can be interpreted within the same framework. 
Near the gelation threshold, the network remains weakly connected and composed of short-lived, weakly associated helices with low constraint density. Under deformation, these dynamic junctions relax and rearrange faster than stress can accumulate along the network, preventing the development of network tension and collective alignment required for strain stiffening. At deeper quenches, the increased density and lifetime of intermolecular associations allow deformation to progressively orient and stretch correlated network segments before yielding, giving rise to the observed stiffening regime. $\iota$-carrageenan remains governed by reversible constraint formation, homogeneous stress redistribution, and progressive network reinforcement over the entire temperature range explored. This interpretation remains consistent with classical descriptions of $\iota$-carrageenan as a weakly aggregated and dynamically connected gel network~\cite{morris1980,robinson1980,geonzon2019}, while extending these structural models toward a mechanical picture in which elasticity, strain stiffening, and reversibility emerge from the progressive formation and deformation of correlated but reversible intermolecular constraints.

\vspace{3mm}
\paragraph{Connection to hierarchical and double-network concepts -- }
The hierarchical organization proposed here shares conceptual similarities with the double-network (DN) hydrogel framework introduced by Gong and coworkers~\cite{gong2003,wang2025}, where distinct structural components contribute separately to elasticity and energy dissipation. In DN hydrogels, a stiff and brittle first network acts as a sacrificial structure that dissipates mechanical energy through internal fracture, while a softer second network maintains macroscopic integrity~\cite{gong2003,sun2020}. 
In $\iota$-carrageenan gels, however, the underlying mechanism is fundamentally different. The two mechanically relevant structural levels do not arise from two distinct molecular constituents, but emerge spontaneously during a single thermoreversible self-assembly of solely $\iota$-carrageenan. A persistent network of intermolecular associations provides the reversible elastic backbone of the gel, while a higher-order organization of helical domains or cooperative intermolecular associations contributes to strain-induced reinforcement. Large deformations selectively disrupt this mechanically adaptive mesoscale organization while preserving sufficient connectivity to recover the small-strain elastic modulus. Thus, rather than relying on irreversible sacrificial fracture, $\iota$-carrageenan exploits a reversible hierarchical architecture in which nonlinear reinforcement can be mechanically remodeled without destroying the underlying elastic network.


\section{Conclusion}
In this work, we have investigated the temperature-dependent linear and nonlinear mechanical response of $\iota$-carrageenan gels formed in KCl solutions. We show that $\iota$-carrageenan forms homogeneous and mechanically reversible networks whose elastic modulus increases continuously with quench depth. The  critical-like evolution of the elastic modulus indicate that network formation proceeds through a progressive increase of intermolecular constraints during cooling, leading to a dynamically connected network capable of reorganizing under deformation.

Beyond linear elasticity, $\iota$-carrageenan gels display a remarkable nonlinear mechanical response. Deep quenches promote pronounced strain stiffening before yielding, accompanied by negative normal forces that reveal the presence of deformation-induced internal stresses within the network. However, large deformations produce only a partial mechanical reorganization: the small-strain elastic modulus is fully recovered, while the strain-stiffening response progressively disappears. This decoupling demonstrates that the structural features responsible for linear elasticity and nonlinear reinforcement are not identical.
By comparing existing descriptions based on filament bending and tensegrity-like stress-bearing networks, we show that neither framework alone can account for the simultaneous recovery of elasticity and loss of strain stiffening. We therefore propose a hierarchical network picture in which a persistent intermolecular network provides the reversible elastic backbone of the gel, while a higher-order organization of helical domains or cooperative intermolecular associations governs strain-induced reinforcement. Large deformations disrupt this mechanically dynamic mesoscale organization without destroying the underlying connectivity of the gel.

More generally, these results show that the mechanics of thermoreversible polysaccharide gels are governed not only by molecular connectivity, but also by the emergence of mechanically adaptive hierarchical structures during network assembly. A major challenge in food hydrocolloid science is to combine firmness, resistance to deformation during processing, and recovery after mechanical stress within a single material. Our results suggest that these functions can be independently tuned by controlling transient hierarchical organizations rather than only the overall network connectivity. The decoupling observed in $\iota$-carrageenan between reversible linear elasticity and irreversible nonlinear reinforcement provides a new framework for designing hydrocolloid networks with tailored texture and processing properties. Extending this framework to $\kappa$-carrageenan, mixed $\iota/\kappa$ formulations, and hybrid carrageenan systems~\cite{moraes2024} may establish hierarchical self-assembly as a general route to independently control the linear and nonlinear mechanics of thermoreversible polysaccharide gels.




\section*{Credit authorship contribution statement}
TG and LH contributed equally to this work.

\section*{Declaration of competing interests}
The authors declare that they have no known competing financial
interests or personal relationships that could have appeared to influence
the work reported in this paper.

\section*{Data availability}
Data is available on reasonable request.

\section*{Acknowledgments}
TG acknowledges the "Région Auvergne-Rhône-Alpes" (AURA grant) in France for funding and the CNRS/Laboratoire de Phyisque (Ens de Lyon) for allowing an extended stay at IPC.
LH acknowledges the "Fundação para a Ciência e Tecnologia" under the framework of Strategic Funding grant UID/CTM/50025/2020 and grant CEECINST/00156/2018.

%

\clearpage

\end{document}